# Comment on "Widely tunable terahertz gas lasers"

Jean-François Lampin[1] and Stefano Barbieri[1]
[1]Institute of Electronics Microelectronics and Nanotechnology, Lille University, CNRS, Centrale Lille, Polytechnique Hauts-de-France, UMR 8520-IEMN, F-59000 Lille, France

**Abstract:** Chevalier *et al.* (Science, 15 November 2019, p. 856-860) report mathematical formulas and a table predicting the threshold and the output power of terahertz molecular lasers based on various molecules. We show that these formulas are not coherent with the simple model used to describe this kind of laser, and that they largely overestimate the conversion efficiency. We suggest an alternative calculation.

Chevalier *et al.* [1] report the demonstration of a terahertz (THz) molecular laser based on $N_2O$ molecules. Two models are used to predict and interpret the results: a simple model (Eq. (1) and (2), derived in part V of the Supplementary Material) and a comprehensive model (part II, III, IV of the Supplementary Material). Experimentally the authors measure a THz power of 10 µW at 0.374 THz, and a slope efficiency of 0.06 mW/W (slope of the linear increase above threshold, Fig. 2B). The authors suggest that the measured output power is underestimated by a factor of four, corresponding to an estimated slope efficiency of about 0.2 mW/W (caption of Fig. 2B). For the same transition, the comprehensive model predicts an output power of 69 µW (slope efficiency:0.4 mW/W). Some experimental difficulties leading to unexpected losses may explain the difference between theory and experiment as mentioned in the article.

On the other hand, in Table 1, the simple model is used to predict the expected output power that can be produced by nine molecules. This model yields a differential efficiency as high as 17.4 mW/W for $N_2O$, i.e. forty times higher than what obtained with the comprehensive model, and three hundred times higher than the experimental value. In this table, the transition considered (0.553 THz) is not the same probed experimentally (0.374 THz) but, as shown in Fig. 1B, this cannot explain the huge difference between the results obtained with the two models. For the 0.553 THz transition, Fig. S5 (or 3C in the main text) shows that the maximum output power predicted by the comprehensive model is close to 0.1 mW, while with the simple model the authors find 4.3 mW (Fig. 1B or Table 1), i.e. a factor of ~43 higher. Even if the simple model is not as accurate as the comprehensive one, there is clearly a problem of coherence between the results obtained with the two models.

To derive the simple model, the authors use a three-level system to describe the gas laser (figure S7). The same kind of model, with 3 or 4 levels, was used in a number of articles (see Refs. [2-4] and references therein). In all these models it is clear that at most only one THz photon can be emitted per IR pump photon absorbed. In some rare cases a cascade process was experimentally observed [5], however in this case two different THz wavelengths are emitted (generally for two different adjustments of the laser cavity) because in a molecule the levels are not equally spaced. Considering a photon conversion efficiency of one (everything is ideal: no losses, parasitic deexcitation, etc...), the maximum power conversion efficiency of the laser is given by:

$$\eta_{MR} = \frac{P_{THz}}{P_{IR}} = \frac{\nu_{THz}}{\nu_{IR}} \qquad (1).$$

More generally this relation is known as the Manley-Rowe limit (MRL) [6], and expresses the maximum power conversion efficiency that can be ideally achieved under the hypothesis of no cascade processes (which is in general the case for non-linear down-conversion generation or optically-pumped lasers). Because the total energy must be conserved, the residual energy must correspond to heat dissipation and/or spontaneous photon emission (in the case of optically pumped lasers), or idler photon generation (in the case of a lossless non-linear parametric process). The MRL is also mentioned by Chevalier *et al.*

In order to check the compatibility of the equations and the predictions of the simple model by Chevalier et al. with the MRL, we take Eq. (S10), and neglect the term $P_{th}$ (which leads to less than 5% error except for CO). We find that if $\alpha_{IR} > 4\alpha_{cell}(R_{cell}/r_0)^2$ (i.e. if $\alpha_{IR} > 6$ m$^{-1}$ with the numerical values given in the caption of Table 1) the predicted efficiency exceeds the MRL. Indeed, from the values given in Table 1 it appears that for $NH_3$ (10.8 m$^{-1}$), OCS (19.6 m$^{-1}$) and $N_2O$ (12.7 m$^{-1}$), more than one THz photon is predicted per IR pump photon (up to 3.3 THz photons / IR photon for OCS). Clearly, this is not physically acceptable, even for an approximated formula, since energy must be conserved (assuming the model of Fig. S7).

As a matter of fact it appears that there is a problem with the approximation done by the authors for the derivation of Eq. (1) and (2) from Ref. [1]. In Eq.(S4) (derivation of Eq. (2)) we read: "The pump power absorbed by the gain medium is approximated as $P_{QCL}(\alpha_{IR}L)$", where $L$ is the laser cavity length. However the expression for the absorbed IR pump power is given by the Beer-Lambert law $P_{QCL}(1-\exp(-\alpha_{IR}d))$, where $d$ is the propagation distance of the pump beam through the gain medium (here, as done by the authors, we neglect possible saturation of the absorption). Therefore the approximation done by the authors is only valid if $\alpha_{IR} d \ll 1$, which is correct only for short distances, and cannot be used in general. For instance, for the N$_2$O transition of Table 1 we find $\alpha_{IR} d$ = 1.9, where $d = L = 0.15$ m is the cavity length used in the paper.

We note that in general, if one wants to extract the maximum power from an optically-pumped laser (i.e. to maximize the efficiency), the absorption of the pump should be almost complete in one or more passes across the cavity. With the author's approximation, $L$ disappears from Eq. S4 because it is also in the denominator. Then the power absorbed per unit length becomes $P_{QCL}\alpha_{IR}$, i.e. the fraction of absorbed pump is larger than one for $\alpha_{IR} > 1/d$, which is clearly not physical (the laser cannot generate pump photons). The same approximation and problem appear in the derivation of Eq. (1) in the main text. This is why the use of these formulas, besides being not physically correct, leads to a strong overestimation of the laser efficiency.

We propose a derivation of Eq. (1) and (2) from Ref. [1], assuming a fraction of absorbed IR pump equal to one (optimum case). Then the pump rate becomes:

$$R_{pump} = \frac{P_{QCL}}{\pi R_{cell}^2 L h \nu_{IR}} \qquad (2),$$

leading to

$$P_{th} = (h^2 \epsilon_0 \nu_{IR}) \left(\frac{u^2 R_{cell}}{\mu_{ij}^2}\right) L \alpha_{cell} \qquad (3),$$

and, finally:

$$P_{THz} = \frac{1}{2} \frac{\nu_{THz}}{\nu_{IR}} T \frac{1}{2L\alpha_{cell}} \left(P_{QCL} - P_{th}\right) \qquad (4),$$

where T is the transmission coefficient of the output coupler. We note that $\alpha_{cell}$ takes into account the ohmic losses of the cavity but also the useful losses, i.e. the transmission of the output coupler (see the end of part II of the Supplementary Material). Then Eq.(4) can be written as

$$P_{THz} = \frac{1}{2} \frac{\nu_{THz}}{\nu_{IR}} \frac{T}{T+A} \left(P_{QCL} - P_{th}\right) \qquad (5),$$

where

$$A = 2L\alpha_{cell} - T \qquad (6).$$

Here, $A$ gives the ohmic losses of the THz waveguide per pass (valid only if $2L\alpha_{cell} \ll 1$). In Ref. [1] the length $L$ of the laser cavity is 0.15 m and the transmission $T$ of the output coupler is 0.04. Two values of $\alpha_{cell}$ are used by the authors: 0.3 m$^{-1}$ for the comprehensive model at 0.374 THz (obtained from a fitting) and 0.06 m$^{-1}$ for the simple model at 0.553 THz (obtained from the theoretical ohmic losses of the TE$_{01}$ mode). In the latter case it seems that the authors forgot to include the transmission of the output coupler because we find $A<0$. By taking it into account, we find $\alpha_{cell}$ = 0.2 m$^{-1}$. With this value, we find that the predicted differential efficiency for N$_2$O (at 0.553THz), obtained with Eq. (4) and (5), is 2.7 mW/W, i.e. six times lower than what predicted by the original simple model in this particular case. Eqs. (3) et (4) can also be used to re-evaluate Table 1, which changes completely the ranking of the molecules.

It should be mentioned that Eq. (5) above corresponds to what was obtained in the past assuming that (i) the pump power is totally absorbed, (ii) the excited state does not absorb the THz photons and (iii) the degeneracy of the two laser levels is the same [2,4]. In the case of a perfect waveguide with no losses ($A = 0$) we find:

$$P_{THz} = \frac{1}{2} \frac{\nu_{THz}}{\nu_{IR}} \left(P_{QCL} - P_{th}\right) \qquad (7)$$

We can see that in this ideal case, if the pump power is sufficiently high to neglect $P_{th}$, the emitted THz power corresponds to half the value of the MRL [7,8] and is considered to be the highest value achievable by a mid-infrared pumped molecular gas THz laser. The ½ factor comes from the differential efficiency term of the 3-level model in the case of equal relaxation rates from levels 2 and 3 [9]. This differential efficiency is equal to 1 if the relaxation from the upper level is suppressed: in this ideal case the MRL can be achieved.

Recently, with the NH$_3$ molecule, we achieved experimentally 55% of the value given by equation (7) (the highest reported to our knowledge), corresponding to a slope efficiency of 10 mW/W [10].

**References**

[1] Chevalier *et al.*, Science, Vol. 366, 856-860 (2019).
[2] D. T. Hodges, Infrared Physics, vol. 18, 375-384 (1978).
[3] J. O. Henningsen and H. G. Jensen, IEEE J. of Quantum Electronics, Vol. QE-11, 248-252 (1975).
[4] J.-M. Lourtioz et al., Revue de Physique Appliquée, vol. 14, 323-330 (1979), in this reference there is an inversion between $\nu_{FIR}$ and $\nu_{IR}$ in the $\eta_{th}$ formula.
[5] T. Y. Chang *et al.*, Applied Physics Letters, vol. 17, 357-358 (1970).
[6] J. M. Manley and H. E. Rowe, Proceedings of the IRE, vol. 44, 904-913 (1956).
[7] E. R. Mueller, Optically pumped terahertz (THz) lasers, in *Wiley Encyclopedia of Electrical and Electronics Engineering*, J. Webster (Ed.), 1996.
[8] A. Pagies *et al.*, Applied Physics Letters Photonics, vol. 1, 031302 (2016).
[9] J. Faist, Quantum cascade lasers, Oxford (2013). Using the notations of chapter 7, it can be seen that when $\tau_{32} \gg \tau_2$ the effective lifetime converges to the upper state lifetime $\tau_3$. The differential efficiency is then $\tau_3/(\tau_2+\tau_3)$ which is equal to ½ if $\tau_3=\tau_2$ and to 1 if $\tau_3 \gg \tau_2$.
[10] J.-F. Lampin *et al.*, Optics Express, vol. 28, 2091 (2020).